\documentstyle[12pt]{article}

\begin{document}
\thispagestyle{empty}
{\hfill  Preprint JINR E2-93-225}\vspace{2.5cm} \\
\vspace*{3cm}
\begin{center}
{\Large\bf  Antibrackets and Supersymmetric Mechanics}\\
\bigskip
\bigskip
\bigskip
\bigskip
{\large Armen Nersessian}
{\footnote{E-MAIL:NERSESS@THEOR.JINRC.DUBNA.SU}}
\bigskip
\vspace*{0.8cm}\\
{\it Laboratory of Theoretical Physics, JINR}\\
{\it Dubna, Head Post Office, P.O.Box 79, 101 000 Moscow, Russia}\\
\end{center}
\bigskip
\bigskip
\bigskip
\bigskip

 \begin{abstract}
Using odd symplectic structure constructed over tangent bundle of the
symplectic manifold, we construct the simple supergeneralization of an
arbitrary Hamiltonian mechanics on it. In the case, if
the initial mechanics defines Killing vector of some Riemannian metric,
corresponding supersymmetric mechanics can be reformulated in the terms
of even symplectic structure on the supermanifold.
\end{abstract}
\vfill
\setcounter{page}0
\renewcommand{\thefootnote}{\arabic{footnote}}
\setcounter{footnote}0
\newpage
\setcounter{equation}0
\section{Introduction}
It is well-known that on supermanifolds ${\cal M}$  the Poisson brackets
of two types can be
defined -- even and odd ones, in correspondence with their Grassmannian
grading~$^1$. That is defined by the expression
\begin{equation}
\{ f, g \}_{\kappa} = \frac{\partial_r f}{\partial z^A}
\Omega^{AB}_{\kappa}(z)
\frac{\partial_l g}{\partial z^B} ,
\label{eq:bloc}
\end{equation}
which  satisfies the  conditions
     \begin{eqnarray}
& &p(\{ f, g \}_{\kappa} )= p(f)+ p(g) + 1   \quad {\rm (grading \quad
condition)} , \nonumber \\
& &\{ f, g \}_\kappa = -(-1)^{(p(f)+\kappa )(p(g)+\kappa )}\{ g, f \}_\kappa
\quad {\rm ( "antisymmetrisity")} ,\label{eq:anti} \\
& & ( -1)^{(p(f)+\kappa )(p(h)+\kappa )}\{ f,\{ g, h \}_{1}\}_1 +
{\rm {cycl. perm. (f, g, h)}} = 0
  \quad{\rm {(Jacobi\quad id.)}},
                  \label{eq:bjac}
\end{eqnarray}
where $z^A$ are the local coordinates on ${\cal M}$,
$ \frac{\partial_r }{\partial z^A}$ and
$\frac{\partial^l}{\partial z^A} $
denote correspondingly the right and the left derivatives,
 $\kappa= 0, 1$ denote correspondingly the even and the odd
Poisson brackets.

Obviously, the even Poisson brackets can be nondegenerate only if
$dim {\cal M}=(2N.M)$, and  the odd  one if $dim {\cal M}=(N.N)$ .

 With nondegenerate Poisson bracket one can associate the
symplectic structure
\begin{equation}
\Omega_{\kappa} =dz^A \Omega_{(\kappa )AB}dz^B ,  \label{eq:symp}
\end{equation}
 where $\Omega_{(\kappa )AB}\Omega^{BC}_{\kappa} =\delta_{A}^{C}$.

Locally nondegenerate even and odd Poisson brackets can be reduced
correspondingly to the forms~$^{2,3}$:
   \begin{equation}
		  \{f,g\}^{\rm can}_0
		     =
			\sum_{i=1}^N \left(
	       \frac{\partial f}{\partial  x^i}
	       \frac{\partial g}{\partial x^{i+N} }
		       -
		\frac{\partial f}{\partial  x^{i+N}}
		\frac{\partial g}{\partial  x^i}
			   \right)
			   +
			  \sum_{\alpha =1}^{M}
	    \epsilon _\alpha \frac{\partial_{r} f}{\partial  \theta_ \alpha}
   \frac{\partial_{l} g }{\partial  \theta_{\alpha}} ,\;\;\epsilon_{\alpha}
=\pm 1 ,
\label{bcan0}   \end{equation}
and
\begin{equation}
		  \{f,g\}^{\rm can}_{1} =
			\sum_{i=1}^{N}\left(
		\frac{\partial_{r} f}{\partial  x^i}
		\frac{\partial_{l} g}{\partial\theta_i}
			 +
              \frac{\partial_{r} f}{\partial  \theta_i}
		\frac{\partial_{l} g}{\partial  x^i}
		   \right).
\label{eq:bcan1}
\end{equation}
The even Poisson brackets are the straightforward generalization of the
ordinary
Poisson brackets on the manifolds. They are widely used in physics for the
description of the
Hamiltonian systems. After quantization such systems described the  theories
 containing both fermionic (real and ghosts) and bosonic degrees of freedom.

The odd Poisson brackets were introduced in field theory by  Batalin
and Vilkovisky~$^4$ (they called them antibrackets, and following thats ,
 we will use the same name) for the formulation of the
covariant (Lagrangian) quantization formalism for the field theories
with arbitrary  constraints (BV-formalism).
BV-formalism and its generalizations~$^5$ are most general methods of the
quantization of the gauge theories.
Recently interest has been aroused in its investigation.
It was stimulated by the  papers~$^6$, which started developing the
background--independent string field theory on the basis of
BV-formalism .

However, the antibrackets can be also used in the direct way -- for the
formulation of the Hamiltonian mechanics with antibrackets ( anti-Hamiltonian
mechanics) :
\begin{equation}
 \frac{dz^A}{dt}=\{z^A , Q\}_{1}, \quad p(Q)=1 . \nonumber
\end{equation}
Leites~$^2$ was the first to point out the possibility of the
formulation of such Hamiltonian mechanics.

However, there were a number of obstacles in the quantization of the
anti-- Hamiltonian mechanics and in the physical interpretation of its
grassmannian
degrees of freedom: the violated  spin--statistics coupling ,
the necessity of
introduction of the odd Planck constant etc.

Possibilities to get over such difficulties were pointed out by
D. V. Volkov et al.~$^{7, 8}$.
Particularly, in~$^8$ it was shown that
one-dimensional Witten's supersymmetric mechanics can be described using
antibrackets, and the role of Hamiltonian in this case plays one of  its
supercharges.

Later the Hamiltonian systems, provided
with  both even and odd Poisson brackets were studied in more detail in~$^9$.

Nevertheless, the absence of the physical examples, where introduction of
the structure  of anti--Hamiltonian mechanics was necessary, or at least
 successful, gave them the status of non-interests systems.

Resently in$^{10, 11}$   the new method of the exact
evaluation of the Hamiltonian path integrals was developed. It is
 based on
the generalization of
the Duistermaat-- Heckman localization formula$^{12}$
{}.
Using this generalization, one can localize the path integral to
the finite-dimensional integral over classical phase space.
This forms the  basis for the conceptually new approach to the invariant
description of supersymmetric
theories.

In$^{13}$ it was shown that it is convenient to use for the description of
this
method the odd symplectic structure, constructed on the supermanifold
associated with the tangent bundle of symplectic manifold, and the
corresponding
Hamiltonian dynamics (anti-Hamiltonian dinamics).

Parallelly, this gives the supersymmetrization method for a  wide class
of the Hamiltonian systems (namely, for the Hamiltonian systems which
define Killing vectors of the some Riemannian metrics on the phase space),
for which there exists a way to go round the noted difficulties connected
with
the quantization.

In the present paper we shall study such Hamiltonian systems and their
supersymmetrization method more closely.

The paper is organized in the following way:

In {\it Section 2} we shall construct the odd symplectic structure and
the corresponding antibrackets on the supermanifold, associated with the
tangent bundle of the initial symplectic manifold. Then we shall define
the natural map of the mechanics on the initial manifold in the supersymmetric
anti-Hamiltonian mechanics, and interpret this supersymmetry in  terms of
the basic manifold.

In {\it Section 3} we shall show that if the initial mechanics defines the
Killing
vector of some Rieman\-nian met\-rics on the phase space, then the
cor\-res\-pon\-ding
su\-per\-sym\-met\-ric anti-Ha\-mil\-to\-nian mecha\-nics can be
re\-for\-mu\-lated
with the
 even Poisson brackets.

\setcounter{equation}0

\section{ Anti-Hamiltonian Systems  and Supersymmetry}

Let  $M$ be the  manifold  with the symplectic structure
\begin{equation}
\omega =\frac{1}{2}\omega_{ij}dx^{i}\wedge dx^{j},
\end{equation}
 and
 \begin{equation}
     \{f(x),g(x)\} = \frac{\partial f}{\partial x^i}\omega^{ij}
\frac{\partial g}{\partial x^j}
\label{eq:beven}\end{equation}
 is the corresponding nondegenerate  Poisson bracket on it.

 The Hamiltonian $H(x)$ defines on it the Hamiltonian mechanics with the
equations of motion:
\begin{equation}
\frac{dx^{i}}{dt}= \{x^{i}, H (x)\}\equiv \xi^{i}.
\label{eq:xi}\end{equation}

It is known that any supermanifold can be associated with some vector
 bundle~$^1$. On the supermanifold which is associated with the cotangent
bundle of any manifold, one can construct the odd symplectic
structure~$^{2}$ ( corresponding Poisson brackets known in
mathematics as Schowten brackets).
Indeed, let  ${\cal M}$ be the supermanifold, associated
with $M$. Then on every map on ${\cal M}$ one can choose the local
coordinates $(x^{i}, \theta_{i})$ ($p(\theta_{i})=p(x^{i})+1=1$),
which are transformed from map to map in accordance with
  \begin{equation}
x^{i} \rightarrow {\tilde x}^{i} ={\tilde x}^{i}(x),\quad
       \theta_{i} \rightarrow {\tilde \theta}_{i} = \sum_{i=1}^N
  \frac{\partial x^{j}}{\partial {\tilde x}^{i}}\theta_{j},
\label{eq:trans}\end{equation}
i.e. $\theta_i$ corresponds to $\frac{\partial}{\partial x^i} $.

Then, obviously, using these coordinates, one can globally define on
${\cal M}$ the antibrackets (\ref{eq:bcan1}).
Let us map the functions on $M$ onto the odd ones on ${\cal M}$:
\begin{equation}
  f(x) \to Q_{f}(z) =\{f(x), F(x,\theta)\}_{1}  ,
\label{eq:map}\end{equation}
where
\begin{equation}
  F(z) = \frac{1}{2}\theta_i \omega^{ij} \theta_j
\label{eq:F0}\end{equation}
 corresponds to the Poisson bracket (\ref{eq:beven}) on $M$.

It is easy to see that
\begin{equation}
 \{ f(x), g(x)\} =\{f(x), Q_{g}(x, \theta )\}_{1} \quad{\rm for}
\quad {\rm any}  \quad f(x), g(x) .
\label{eq:cons}
\end{equation}
Then, the map (\ref{eq:map}) puts the Hamiltonian mechanics $(H, \omega , M)$
into the anti-Hamiltonian mechanics
$\left ( Q,\Omega_{1}, {\cal M}\right )$, where
\begin{equation}
Q\equiv Q_{H}=\{H, F\}_{1}.
\label{eq:q}\end{equation}
This mechanics  is supersymmetric.
Indeed, it is easy to see that the functions $H, F, Q$ form the simplest
superalgebra:
\begin{eqnarray}
&Q=\{H, F\}_{1},&\quad \{H, H\}_{1}=0,\quad \{F, F\}_{1}=0 ,
\label{eq:sualg}\\
&\{Q, Q\}_{1}=0,&\quad \{H, Q\}_{1} =\{F, Q\}_{1}=0  ,\nonumber
\end{eqnarray}
or, equivalently,
\begin{eqnarray}
 && \{H\pm F, H \pm F\}_{1} =\pm 2Q ,\\
 && \{H + F , H - F\}_{1} = \{H\pm F, Q \}_{1} = \{Q, Q \}_{1} = 0 .
              \nonumber
\label{eq:sualg2} \end{eqnarray}
The last equation  in (\ref{eq:sualg}) corresponds to the Jacobi
identity for  (\ref{eq:beven}).

For the interpretation of this superalgebra note that in (\ref{eq:map})
$Q_{f}$ is transformed as $df$. Correspondingly,
\begin{equation}
\theta^{i}=\{x^i , F\}_{1}=\omega^{ij}\theta_j
\label{eq:theta}\end{equation}
 can be interpreted as the basic 1-forms $dx^i$.

Then, one can pass from the desctiption in terms of the coordinates
$(x^i , \theta_i )$ to that in terms of the coordinates $(x^i , \theta^i )$.

Obviously, any function $f(x^i ,\theta^i )$ can be interpreted in terms of
differential forms on $M$.

In terms of $(x^i , \theta^i )$ antibrackets  (\ref{eq:bcan1})
on ${\cal M}$ take the form:
\begin{equation}
\{x^i, x^j\}_{1}=0,\quad \{x^i, \theta^j\}_{1}=\omega^{ij},
\quad\{\theta^i, \theta^j\}_{1}=
\frac{\partial\omega^{ij}}{\partial x^k}\theta^k
\label{eq:bxt}\end{equation}
where $\omega^{ij}$ is the matrix of the even Poisson bracket
(\ref{eq:beven})
on $M$.

The corresponding odd symplectic structure in coordinates
$z^A =(x^i ,\theta^i )$
 takes the form
\begin{equation}
\Omega_{1} =\frac{1}{2}\omega_{ij}dx^i\wedge d\theta^j
+\frac{1}{2}\omega_{ij,k}\theta^k dx^i\wedge dx^j
\label{eq:osym}\end{equation}

The equations of motion of the anti-Hamiltonian mechanics
$\left ( Q, \Omega_{1}, {\cal M}\right )$ are :
\begin{equation}
\frac{d x^i}{dt}=\{x^{i} , Q\}_{1}=\xi^i \quad\frac{d\theta^i }{dt} =
\{\theta^{i} , Q\}_{1} =
\frac{\partial \xi^i}{\partial x^j}\theta^j.
\label{eq:motion}\end{equation}

The following correspondence is obvious
\begin{eqnarray}
&& \{H,\quad\}_{1}=\xi^i \frac{\partial}{\partial \theta^i}
 \rightarrow \imath _{H} -{\rm operator \quad of\quad inner\quad
product\quad on}\quad \xi^i ;\nonumber\\
&&\{F,\quad\}_{1}= \theta^i
\frac{\partial}{\partial x^i}   \rightarrow  d -{\rm operator\quad of\quad
 exterior\quad differentiation} ;\nonumber\\
&&\{ Q,\quad \}_{1}=\xi^i
 \frac{\partial}{\partial x^i} + \xi_{,k}^{i}\theta^k \frac{\partial}{\partial
 \theta^i} \rightarrow  {\cal L}_{H} -{\rm Lie\quad derivative\quad
along}\quad \xi^i .\nonumber\\
\label{eq:corr}\end{eqnarray}
Then, using the Jacobi
identity (\ref{eq:bjac}), we show
\begin{equation} \{ H, F\}_{1} =Q
\rightarrow d\imath_{H} +\imath_{H}d = {\cal L}_{H} - {\rm homotopy \quad
formula}\nonumber
\end{equation}
As we see, the supercharge  $H+ F$ which
defines the supersymmetry transformation , corresponds to the  operator of
the equivariant differentiation $d_{H}= d+\imath_{H}$.

Therefore the presented anti-Hamiltonian mechanics is the natural canditate for
the
description of equivariant localization$^{13}$.

However, and we noted that in the Introduction, there are many obstacles
in the quantization and interpretation of the anti-Hamiltonian mechanics.
In the next Section we shall show that some additional
assumption about the anti-Hamiltonian mechanics (\ref{eq:motion})
allows reformulate it with the even Poisson brackets and to go round
this obstacles.

\setcounter{equation}0
\section{Reformulation with Even Poisson Brackets}

In the previous Section we saw, that defining the map (\ref{eq:map}), we
go from the arbitrary Hamiltonian mechanics (\ref{eq:xi}) to the
supersymmetric
one (\ref{eq:motion}).

In this Section we shall consider the special case of the mechanics
(\ref{eq:xi}), when it defines the Killing vector for some Riemannian
 mertic on the phase space.

In this case the corresponding supersymmetrization (\ref{eq:motion}) can be
reformulated in terms of the {\it even} symplectic structure on the
phase {\it superspace}.

Let us assume that the manifold $M$ is provided with both the symplectic
structure
$\omega_{\alpha}$ and the Riemannian one $g_{ij}$.

Let  the local 1-form $A_{\alpha}=A_{(\alpha) i}dx^{i}$ define on $M$
this symplectic structure:
 $$dA_{\alpha} =\omega_{\alpha}. $$

Then  consider on ${\cal M}$ the following local 1-form:
\begin{equation}
{\cal A}_{\alpha}=
A_{(\alpha)i}dx^{i} + \theta^{i}g_{ij}D\theta^j \label{eq:A},
\end{equation}
where $D\theta^{i} = d\theta^{i} + \Gamma^{i}_{kl}\theta^{k}dx^{l}$
and $\Gamma^{i}_{kl}$ -- the Cristoffel symbols for metrics $g_{ij}$ on $M$.

The exterior differential of this 1-form (globally) define
on ${\cal M}$ the even symplectic structures:
\begin{equation}
{\Omega}_{\alpha}=d{\cal A}_{\alpha}=
\frac{1}{2} (\omega_{(\alpha ) ij}
+ R_{ijkl}\theta^{k}\theta^{l})dx^{i}\wedge dx^{j} +
g_{ij}D\theta^{i}\wedge D\theta^{j},
\label{eq:Oalpha}\end{equation}
where
$R_{ijkl} $ -- the curvature tensor on $M$.

 The Poisson brackets, which correspond to this structure, are :
\begin{equation}
\{f(z) , g(z)\}_{\alpha}= \nabla_{i}f(z)(\omega_{(\alpha ) ij}
+ R_{ijkl}\theta^{k}\theta^{l})^{-1}\nabla_{j}g(z) +
\frac{\partial_{r}f(z)}{\partial \theta^i}
g^{ij}\frac{\partial_{l}g(z)}{\partial \theta^j},
\end{equation}
where $g^{ik}g_{kj}=\delta^{i}_{j}$,
$$\nabla_{i}=\frac{\partial}{\partial x^i} -
\Gamma^{k}_{ij}\theta^{j}\frac{\partial_{l}}{\partial \theta^{k}}.$$

Now let us assume that  the Riemannian metric $g_{ij}$ on $M$
 Lie-derived with $\xi^{i}$ (\ref{eq:xi}):
\begin{equation}
{\cal L}_{H}g =0 \Leftrightarrow \xi^{k}_{,i}g_{kj} + g_{ik}\xi^{k}_{,j} +
     g_{ij,k}\xi^{k} =0 .
\label{eq:metric}\end{equation}

Then  the odd function
\begin{equation}
{\tilde Q} = \xi^i g_{ij}\theta^j
\label{eq:gauge} \end{equation}
on ${\cal M}$ is the motion integral of the anti-Hamiltonian mechanics:
\begin{equation}
{\cal L}_{H} g = 0 \rightarrow \{Q, {\tilde Q}\}_{1}=0 .
\label{eq:killing}\end{equation}

The functions $F$ and $H$ commute with ${\tilde Q}$ in the
following way:
\begin{equation}
   \{F, {\tilde Q} \}_{1}= - F_{2} ,\quad \{H, {\tilde Q}\}_{1} = H_{2}   ,
\label{eq:FQ,HQ}\end{equation}
where
\begin{equation}
 H_{2}= \xi^{i}g_{ij}\xi^{j},\quad
F_2 =\frac{1}{2}\theta^{i}\omega_{(2)ij}\theta^j,
\quad \omega_{(2)ij}=\frac{\partial ( g_{ik}\xi^{k}_{H})}{\partial x^j} -
\frac{\partial ( g_{jk}\xi^{k}_{H})}{\partial x^i}.
\label{eq:F1H1} \end{equation}
Let us assume that $\det{\omega_{(2)ij}}\neq 0$.

Then,
the mechanics $(H_{2}, \omega_{(2)ij}dx^{i}\wedge dx^{j}, M)$
   and $(H,\omega, M)$
define the bi-Hamiltonian structure on $M$ (it was first pointed out
 first in$^{12}$):
\begin{equation}
\xi^{i}=\omega^{ij}\frac{\partial H}{\partial x^j}
=\omega_{(2)}^{ij}\frac{\partial H_{2}}{\partial x^j}.
\end{equation}
  It is easy
 to see that $({\cal H}_{\alpha}= H_{\alpha} +F_{2},
{\Omega}_{\alpha}, {\cal M})$
 and $(Q, {\Omega}_{1}, {\cal M})$
 define the same  Hamiltonian
 dynamics (\ref{eq:motion}) on ${\cal M}$.

There  $\alpha =0, 2$ and ${\Omega}_{\alpha}$ is defined by the
(\ref{eq:Oalpha}), $\omega_{0}\equiv \omega$, $H_{0}\equiv H$.

In other words, we pro\-vide the su\-per\-sym\-met\-ric
anti-Hamil\-to\-ni\-an mecha\-nics
(\ref{eq:motion}) with the {\it even} {\it Hamiltonian} {\it
structure}
in the case, that the initial mechanics $( H, \omega, M)$ define the
Killing vector for the some Riemannian metric on the $M$.

Using  the description with it we can go round the obstacles
connected with the difficultes
in the quantization and
interpretation  of  anti-Hamiltonian systems.

However, on the level of classical description the use of
antibrackets make its more simple and transparent.

It is obvious that this dynamics has at least two supercharges.

We have got the simple supersymmetrization method
for the  Hamiltonian systems,
which is defined on the symplectic manifolds,
provided with the Riemannian structures,
Lie-derived with it.

This class includes the integrable systems
 on the orbits of the coadjoint representation of semisimple Lie groups
and therefore, really all integrable systems of classical mechanics.

Example (1D supersymmetric Witten mechanics) of dynamics
 with even and odd Hamiltonian  structures
was considered
at first in$^{8}$ (see also$^{9}$).

 \bigskip
{\Large\bf References}
 \bigskip
\begin{enumerate}
\item F. A. Berezin -- Introduction to Superanalysis.,
      D. Reidel, Dordrecht, 1986

T. Voronov -- Geometric Integration Theory on Supermanifolds. Sov. Sci. Rev. C,
 Math.Phys., v.9, 1992
\item D. A. Leites  -- Dokl. Akad. Nauk SSSR, {\bf 236} (1977), 804
\item V. N. Shander -- Dokl. Akad. Nauk. Bulgaria {\bf  },   (1983)
\item I. A. Batalin , G. A. Vilkovisky -- Phys.Lett., {\bf 102B} (1981), 27;
 Nucl.Phys., {\bf  B234} (1984), 106
\item I. A. Batalin, P. M. Lavrov, I. V. Tyutin --J. Math. Phys., {\bf 31}
(1990) 1487; {\it ibid.} {\bf 32} (1991), 532; {\it ibid.} {\bf 32} (1991),
2513

I. A. Batalin, I. V. Tyutin -- Preprint FIAN/TD/18-92,
 Int. J. Mod. Phys. A , {\bf } (1993)
\item E. Witten-- Mod. Phys. Lett. A, {\bf 5} (1990), 487; Preprint
IASSNS-HEP- 92/53 ; Preprint IASSNS-HEP- 92/63
\item D. V. Volkov -- JETP Lett., {\bf 38} (1983), 508

D. V. Volkov , V. A. Soroka , V. I. Tkach  -- Sov. J. Nucl. Phys.,
{\bf 46} (1987) , 110
\item D. V. Volkov , V. A. Soroka , A. I. Pashnev , V. I.Tkach  -- JETP Lett.,
  {\bf 44} (1986), 55
\item O. M. Khudaverdian , A. P. Nersessian -- Preprint  YERPHI-1031(81)-1987;
     J. Math. Phys., {\bf 32} (1991), 1938 ; Preprint JINR E2-92-411,
J. Math. Phys.(to appear)

O. M. Khudaverdian  -- J. Math. Phys., {\bf 32} (1991), 1934 ;

A. P. Nersessian  -- Theor. Math. Phys.,{\bf 96} (1993), No. 1 (in press)
\item M. Blau, E. Keski- Vakkuri , A. J. Niemi
-- Phys.Lett., {\bf 246B} (1990), 92;

A. Hietaki, A. Yu. Morozov, A. J. Niemi, K. Palo -- Phys. Lett. {\bf B263}
(1991), 417

A. Yu. Morozov, A. J. Niemi, K. Palo -- Phys. Lett {\bf B271} (1991), 365;
 Nucl. Phys. {\bf B377} (1992), 295


A. J. Niemi, O. Tirkkonen  -- Phys.Lett., {\bf 293B} (1992), 339;

 E. Witten  -- Preprint IASSNS-HEP-92/15
\item A. J.  Niemi, O. Tirkkonen -- Preprint UU-ITP 3/93

\item J. J. Duistermaat , G. J. Heckman -- Inv. Math. {\bf 69} (1982), 259;
{\it ibid.}{\bf 72} (1983), 153

 M. F. Atiah , R. Bott  -- Topology, {\bf 23}, No. 1 (1984), 1
\item A.P. Nersessian -- JETP Lett., {\bf 58} (1993), No. 1 (in press);
\end{enumerate}

\end{document}